\documentclass[reqno]{amsart}
\usepackage{amssymb,latexsym,amsfonts,amsmath}
\usepackage{graphicx}
\usepackage{paralist}
\usepackage{etoolbox}
\usepackage{cite}
\usepackage{eso-pic}
\usepackage{epsfig}
\usepackage{mathtools}
\usepackage{tikz}
\usepackage{epstopdf}
\usepackage{graphicx}
\usepackage{epsfig}
\usepackage{mathtools}
\usepackage{amsfonts}
\usepackage{lineno}
\usepackage{tikz}
\usetikzlibrary{positioning,shapes}
\usetikzlibrary{arrows}
\usepackage[justification=centering]{caption}
\allowdisplaybreaks
\AtBeginEnvironment{thebibliography}{\linespread{0.91}\selectfont}
\newtheorem{theorem}{Theorem}
\newtheorem{definition}[theorem]{Definition}

\newtheorem{assumption}[theorem]{Assumption}
\newtheorem{proposition}[theorem]{Proposition}
\newtheorem{remark}[theorem]{Remark}

\newcommand{\TF}{\mathcal{F}}
\newcommand{\norm}[1]{\left\Vert #1\right\Vert}

\newcommand{\op}{{\mathcal H}}
\def\field#1{\mathbb #1}%
\def\R{\field{R}}%
\def\N{\field{N}}%
\def\Z{\field{Z}}%

\newcommand{\SF}{\mathcal{S}}
\newcommand{\Let}{:=}
\newcommand{\KK}{\mathcal{K}_{\infty}}
\newcommand{\intcc}[1]{\ensuremath{{\left[#1\right]}}}

\usepackage{color}

\usepackage{xspace}

\renewcommand{\emptyset}{{\varnothing}}
\usepackage{enumerate}

\begin{document}
\title[Compositional Abstractions of Interconnected Discrete-Time Switched Systems]{Compositional Abstractions of Interconnected Discrete-Time Switched Systems}
	\author{Abdalla Swikir$^1$}
\author{Majid Zamani$^2$$^3$}
\address{$^1$Hybrid Control Systems Group, Technical University of Munich, Germany.}
\address{$^2$Computer Science Department, University of Colorado Boulder, USA. }
\address{$^3$ Computer Science Department, Ludwig Maximilian University of Munich, Germany.}
\email{abdalla.swikir@tum.de, majid.zamani@colorado.edu}
\maketitle

	\begin{abstract}                          
		In this paper, we introduce a compositional method for the construction of finite abstractions of
		interconnected discrete-time switched systems. Particularly, we use a notion of so-called alternating simulation function as a relation between each switched subsystem and its finite abstraction. Based on some small-gain type conditions, we use those alternating simulation functions to construct compositionally an overall alternating simulation function as a relation between an interconnection of finite abstractions and that of switched subsystems. This overall alternating simulation function allows one to quantify the mismatch between the output behavior of the interconnection of switched subsystems and that of their
		finite abstractions. 
		Additionally, we provide an approach to construct finite abstractions together with their corresponding alternating simulation functions for discrete-time switched subsystems under standard assumptions ensuring incremental input-to-state stability of a switched subsystem.
		Finally, we apply our results to a model of road traffic by constructing compositionally a finite abstraction of the network containing $50$ cells of $1000$ meters each. We use the constructed finite abstractions as substitutes to design controllers compositionally keeping the density of traffic lower than $30$ vehicles per cell.
	\end{abstract}
	
	\section{Introduction}
	
	Switched systems serve as an important modeling framework accurately describing several engineering systems
	in which physical processes have various operational modes \cite{liberzon}. Despite considerable number of studies that have been conducted regarding stability of switched systems, the fast grow
	in computational technology requires us to make same progress with
	respect to more sophisticated objectives such as those expressed as linear temporal logic (LTL) formulae \cite{Katoen}. One particular technique to address complex objectives is based on the construction of
	finite abstractions (a.k.a. symbolic models) of switched systems. In the finite abstractions, each abstract state represents a collection of continuous states of the switched system. 
	Since finite abstractions are finite, one can algorithmically solves controller synthesis problems by resorting to automata-theoretic approaches \cite{MalerPnueliSifakis95,Thomas95}. In general, there exist two types of finite abstractions: \emph{sound} ones whose behaviors (approximately) contain those of the concrete systems and \emph{complete} ones whose behaviors are (approximately) equivalent to those of the concrete systems \cite{Tabu}.
	
	In recent years, there have been several results on the construction of complete finite abstractions of switched systems.
	The work by \cite{Girard} provides a finite abstraction that is related to the original incrementally stable switched system by establishing an approximate bisimulation relation between them. 
	Recently, the result in \cite{Girard} has been extended to the case of multi-rate symbolic models in \cite{SAOUD}, multi-scale symbolic models in \cite{Gossler}, and to switched systems with aperiodic time sampling in \cite{KADER}.
	
	All the proposed results in \cite{Girard,Corronc,Gossler,SAOUD,KADER} take a monolithic view of switched systems when abstracting the entire system. 
	However, the computational complexity of constructing finite abstractions scales exponentially with the number of state variables in the concrete switched system. Hence, the construction of finite abstractions for large-scale interconnected switched systems is mostly a complex task from a computational point of view. A convenient method to cope with this challenge is to first construct finite abstractions of the switched subsystems individually and then establish a compositional scheme that allows to construct a finite abstraction of the overall network using those individual finite abstractions.
	
	In the past few years, several results have used the compositional framework for constructing complete finite abstractions of networks of control subsystems.
	Based on the notion of interconnection-compatible approximate bisimulation relation, \cite{Tazaki2008} provides networks of finite abstractions
	that approximate networks of stabilizable linear control systems. This work was extended in \cite{7403879} to networks of incrementally input-to-state stable nonlinear control
	systems using the notion of approximate bisimulation relation. The work in \cite{Majumdar} introduces a new system relation, called approximate disturbance
	bisimulation relation, as the basis for the compositional construction of finite abstractions. The results in \cite{arxiv, swikir} provide techniques to construct  compositionally finite abstractions of networks of nonlinear control systems using dissipativity and general small-gain type conditions, respectively. There are also other results in the literature \cite{meyer,omar,Kim} which provide sound finite abstractions of interconnected systems, compositionally, without requiring any stability property or condition on the gains of subsystems. Unfortunately, non of the compositional results in \cite{Tazaki2008,7403879,Majumdar,arxiv,swikir,meyer,omar,Kim} provide a compositional framework for the construction of finite abstractions for interconnected switched systems.
	
	The main contribution of this work is to provide for the first time a compositional methodology for the construction of finite abstractions of interconnected switched systems. The proposed approach leverages sufficient small-gain type conditions to establish the compositionality results which rely on the existence of alternating simulation functions as relations between switched subsystems and their finite abstractions. In particular, based on some small-gain type conditions, we use those alternating simulation functions to construct compositionally an overall alternating simulation function as a relation between an interconnection of finite abstractions and that of original switched subsystems. The existence of such an overall alternating simulation function enables one to quantify the mismatch between the output behavior of the interconnection of switched subsystems and that of their
	finite abstractions. Furthermore, under standard assumptions ensuring incremental input-to-state stability of a switched system (i.e., existence
	of a common incremental input-to-state Lyapunov function, or multiple incremental input-to-state Lyapunov functions with dwell-time), we show that one can construct finite
	abstractions of switched systems in general nonlinear settings.
	Finally, we apply our results to a model of road traffic by constructing compositionally a finite abstraction of a network containing $50$ cells of $1000$ meters each. We use the constructed finite abstractions as substitutes to design controllers compositionally maintaining the density of traffic lower than $30$ vehicles per cell. Notation
	and some technical notions used in the sequel are reported in the Appendix.
	\section{Preliminaries}\label{1:II}
	
	\subsection{Discrete-Time Switched Systems} 
	In this paper we study discrete-time switched systems of the following form.
	\begin{definition}\label{def:sys1}
		A discrete-time switched system $\Sigma$ is defined by the tuple $\Sigma=(\mathbb X,P,\mathbb W,F,\mathbb Y,h)$,
		where 
		\begin{itemize}
			\item 	$\mathbb X, \mathbb W,$ and $\mathbb Y$ are the state set, internal input set, and output set, respectively, and are assumed to be subsets of normed vector spaces with appropriate finite dimensions; 
			\item $P=\{1\cdots,m\}$ is the finite set of modes;
			\item $F=\{f_1,\cdots,f_m\}$ is a collection of set-valued maps $f_p: \mathbb X\times \mathbb W\rightrightarrows\mathbb X $ for all $p\in P$;
			\item $h: \mathbb X \rightarrow \mathbb Y $ is the output map.
		\end{itemize} 
		The discrete-time switched system $\Sigma $ is described by difference inclusions of the form
		\begin{small}
			\begin{align}\label{eq:2}
			\Sigma:\left\{
			\begin{array}{rl}
			\mathbf{x}(k+1)&\!\!\!\!\in f_{\mathsf{p}(k)}(\mathbf{x}(k),\omega(k)),\\
			\mathbf{y}(k)&\!\!\!\!=h(\mathbf{x}(k)),
			\end{array}
			\right.
			\end{align}
		\end{small}
		where $\mathbf{x}:\mathbb{N}\rightarrow \mathbb X $, $\mathbf{y}:\mathbb{N}\rightarrow \mathbb Y$, $\mathsf{p}:\mathbb{N}\rightarrow P$, and $\omega:\mathbb{N}\rightarrow \mathbb W$ are the state signal, output signal, switching signal, and internal input signal, respectively.
		We denote by $\Sigma_{p}$ system \eqref{eq:2} with constant switching signal $\mathsf{p}(k)=p\in P ~\forall k\in \N$.
		We use $\mathbf{X}_{x_0,\overline{p},\overline{\omega}}$ and $\mathbf{Y}_{x_0,\overline{p},\overline{\omega}}$ to denote the sets of infinite state and output runs of $\Sigma$, respectively, associated with infinite switching sequence $\overline{p}=\{p_0,p_1,\ldots\}$, infinite internal input sequence $\overline{\omega}=\{w_0,w_1,\ldots\}$, and initial state $x_0\in \mathbb{X}$.
		
		Let $\phi_k, k \in \N_{\ge1}, $ denote the time when the $k$-th switching instant occurs and
		define $\Phi := \{\phi_k : k \in \N_{\ge1}\}$ as the set of switching instants.
		We assume that signal $\mathsf{p}$ satisfies a dwell-time condition \cite{Morse} (i.e. there exists $k_d \in \N_{\ge1}$, called the
		dwell-time, such that for all consecutive switching time instants $\phi_k,\phi_{k+1} \in \Phi$, $\phi_{k+1}-\phi_{k}\geq k_d$). 
		
		System $\Sigma$ is called deterministic if $|f_p(x,w)|\leq1$ $ \forall x\in \mathbb X, \forall p\in P, \forall w \in \mathbb W$, and non-deterministic otherwise. System $\Sigma$ is called blocking if $\exists x\in \mathbb X, \forall p\in P, \forall w \in \mathbb W $ where $|f_p(x,w)|=0$ and non-blocking if $|f_p(x,w)|\neq 0$ $ \forall x\in \mathbb X, \exists p\in  P, \exists w \in \mathbb W$.  System $\Sigma$ is called finite if $\mathbb X$ and $\mathbb W$ are finite sets and infinite otherwise. In this paper, we only deal with non-blocking systems.
	\end{definition}	
	\section{Transition Systems and Alternating Simulation Functions}\label{I}
	In this section, we introduce a notion of so-called transition systems to provide an alternative description of switched systems that can be later directly related to their finite abstractions.
	\begin{definition}\label{tsm} Given a discrete-time switched system $\Sigma=(\mathbb X,P,\mathbb W,F,\mathbb Y,h)$, we define the associated transition system $T(\Sigma)=(X,U,W,\TF,Y,\op)$ 
		where:
		\begin{itemize}
			\item  $X=\mathbb X\times P\times \{0,\cdots,k_d-1\}$ is the state set; 
			\item $U=P$ is the external input set;
			\item $W=\mathbb{W}$ is the internal input set;
			\item $\TF$ is the transition function given by $(x',p',l')\in \TF((x,p,l),u,w)$ if and only if  $x'\in f_p(x,w),u=p$ and the following scenarios hold:
			\begin{itemize} 
				\item$l<k_d-1$, $p'=p$ and $l'=l+1$: switching is not allowed because the time elapsed since
				the latest switch is strictly smaller than the dwell time;
				\item $l=k_d-1$, $p'=p$ and $l'=k_d-1$: switching is allowed but no switch occurs;
				\item $l=k_d-1$, $p'\neq p$ and $l'=0$: switching
				is allowed and a switch occurs;
			\end{itemize}
			\item $Y=\mathbb{Y}$ is the output set;
			\item $\mathcal{H}:X\rightarrow Y$ is the output map defined as $\mathcal{H}(x,p,l)=h(x)$.
		\end{itemize}
	\end{definition}
	We use $T(\mathbf{X})_{z_0,\overline{u},\overline{\omega}}$ and $T(\mathbf{Y})_{z_0,\overline{u},\overline{\omega}}$ to denote the sets of infinite state and output runs of $T(\Sigma)$, respectively,  associated with infinite external input sequence $\overline{u}=\{u_0,u_1,\ldots\}$, infinite internal input sequence $\overline{\omega}=\{w_0,w_1,\ldots\}$, and initial state $z_0=(x_0,p_0,l_0)\in X$, where $u_0=p_0$ and $l_0=0$.
	
	In the next proposition, we show that sets $\mathbf{Y}_{x_0,\overline{p},\overline{\omega}}$ and $T(\mathbf{Y})_{z_0,\overline{u},\overline{\omega}}$, where $\overline{p}\!=\!\overline{u}$ and $z_0\!\!=\!\!(x_0,p_0,0)$, are equivalent.
		\begin{proposition}\label{traj}
			Consider $\Sigma$, $T(\Sigma)$, $\overline{p}=\{p_0,p_1,\ldots\}=\overline{u}$, $\overline{\omega}=\{w_0,,w_1,\ldots\}$, and $x_0\in \mathbb{X}$. Then, $\mathbf{Y}_{x_0,\overline{p},\overline{\omega}}=T(\mathbf{Y})_{z_0,\overline{u},\overline{\omega}}$, 
			where $z_0=(x_0,p_0,0)$.
	\end{proposition}
	The proof is straightforward and omitted here due to lack of space.
	
	From now on, we use $\Sigma$ and $T(\Sigma)$ interchangeably.
	
	In the following, we introduce a notion of so-called alternating simulation functions, inspired by Definition 1 in \cite{Girard2009566}, which quantitatively relates transition systems with internal inputs.
	\begin{definition}\label{sf} 
		Consider $T(\Sigma)=(X,U,W,\TF,Y,\op)$  and $\hat{T}(\hat{\Sigma})=(\hat{X},\hat{U},\hat{W},\hat{\TF},\hat{Y},\hat{\op})$ where $\hat W\subseteq W$ and $\hat Y\subseteq Y$. A function $ \mathcal{S}:X\times \hat X \to \mathbb{R}_{\geq0} $ is called an alternating simulation function from $\hat{T}(\hat{\Sigma})$ to $T(\Sigma)$ if $\forall (x,p,l)\in X$ and $\forall (\hat{x},p,l)\in\hat{X}$,  one has
			\begin{align}\label{sf1}
			\alpha (\Vert \op(x,p,l)\!-\hat{\op}(\hat{x},p,l)\Vert ) \!\leq\! \SF(\!(x,p,l),(\hat{x},p,l)\!),
			\end{align}
		$\!\!$and $\forall (x,p,l)\in X$ and $\forall (\hat{x},p,l)\in\hat{X}$, $\forall \hat u\in\hat{U}$, $\forall w\in W$, $\forall \hat w\in\hat{W}$, $\forall (x',p',l')\in\TF((x,p,l),\hat{u},w)$ $\exists~ (\hat{x}',p',l')\in\hat{\TF}((\hat{x},p,l),\hat{u},\hat{w})$ such that one gets
			\begin{align}\label{sf2}
			\SF((x',p',l'),&(\hat{x}',p',l'))\leq \max\{\sigma \SF((x,p,l),(\hat{x},p,l)),\varrho(\Vert w- \hat{w}\Vert ),\varepsilon\},
			\end{align}
		$\!\!$for some $\alpha,\varrho \in \mathcal{K}_{\infty}$,  $0<\sigma<1$, and $\varepsilon\in \mathbb{R}_{\geq 0}$.
	\end{definition}
	If $\Sigma$ does not have internal inputs, which is the case for interconnected systems (cf. Definition \ref{def:5}), Definition \ref{def:sys1} reduces to the tuple $\Sigma=(\mathbb X,P,F,\mathbb Y,H)$, the set-valued map $f_p$ becomes $f_p:\mathbb X\rightrightarrows\mathbb X$, and \eqref{eq:2} reduces to:
		\begin{align}\label{eq:3}
		\Sigma:\left\{
		\begin{array}{rl}
		\mathbf{x}(k+1)\!\!\!\!&\in f_{\mathsf{p}(k)}(\mathbf{x}(k)),\\
		\mathbf{y}(k)\!\!\!\!&=h(\mathbf{x}(k)).
		\end{array}
		\right.
		\end{align}
	$\!\!$Correspondingly, Definition \ref{tsm} reduces to tuple $T(\Sigma)=(X,U,\TF,Y,\op)$, and the transition function $\TF$ is given by $(x',p',l')\in \TF((x,p,l),u)$ if and only if $ x'\in f_p(x),u=p$ and the following scenarios hold:
	\begin{itemize}
		\item $l<k_d-1$, $p'=p$ and $l'=l+1$;
		\item $l=k_d-1$, $p'=p$ and $l'=k_d-1$;
		\item $l=k_d-1$, $p'\neq p$ and $l'=0$.
	\end{itemize}
	Moreover, Definition \ref{sf} reduces to the following.
	\begin{definition}\label{sfg}
		Consider $T(\Sigma)=(X,U,\TF,Y,\op)$  and $\hat{T}(\hat{\Sigma})=(\hat{X},\hat{U},\hat{\TF},\hat{Y},\hat{\op})$ where $\hat Y\subseteq Y$. A function $ \tilde{\SF}:X\times \hat X \to \mathbb{R}_{\geq0} $ is called an alternating simulation function from $\hat{T}(\hat{\Sigma})$ to $T(\Sigma)$ if $\forall (x,p,l)\in X$ and $\forall (\hat{x},p,l)\in\hat{X}$,  one has
			\begin{align}\label{sfg1}
			\tilde{\alpha} (\Vert \op(x,p,l)-\hat{\op}(\hat{x},p,l)\Vert ) \!\leq\! \tilde{\SF}((x,p,l),(\hat{x},p,l)),
			\end{align}
	
		and $\forall (x,p,l)\in X$ and $\forall (\hat{x},p,l)\in\hat{X}$, $\forall \hat u\in\hat{U}$, $\forall (x',p',l')\in\TF((x,p,l),\hat{u})$ $\exists~ (\hat{x}',p',l')\in\hat{\TF}((\hat{x},p,l),\hat{u})$ such that one gets
			\begin{align}\label{sfg2}
			\tilde{\SF}(\!(x',p',l'),(\hat{x}',p',l')\!)
			\!\leq\! \max\{\!\tilde{\sigma} \tilde{\SF}(\!(x,p,l),(\hat{x},p,l)\!),\tilde{\varepsilon}\}\!,
			\end{align}
		$\!\!$for some $\tilde{\alpha}\in \mathcal{K}_{\infty}$,  $0<\tilde{\sigma}<1$, and $\tilde{\varepsilon}\in \mathbb{R}_{\geq 0}$.
	\end{definition}
	The next result shows that the existence of an alternating simulation function for transition systems without internal inputs implies the existence of an approximate alternating simulation relation between them as defined in \cite{Tabu}.	
	\begin{proposition}\label{error}
		Consider $T(\Sigma)=(X,U,\TF,Y,\op)$ and $\hat{T}(\hat{\Sigma})=(\hat{X},\hat{U},\hat{\TF},\hat{Y},\hat{\op})$ where $\hat Y\subseteq Y$. Assume $ \tilde{\SF}$ is an alternating simulation function from $\hat{T}(\hat{\Sigma})$ to ${T}(\Sigma)$ as in Definition \ref{sfg}. Then, relation $R\subseteq X\times \hat{X}$ defined by $$R\!=\!\left\{\!((x,p,l),\!(\hat{x},p,l))\!\in\! {X}\!\times\! \hat{X}|\tilde{\SF}((x,p,l),(\hat{x},p,l))\!\leq \!\tilde{\varepsilon}\!\right\}$$ is an $\hat{\varepsilon}$-approximate alternating simulation relation, defined in \cite{Tabu}, from $\hat{T}(\hat{\Sigma})$ to ${T}(\Sigma)$ with $\hat{\varepsilon}=\tilde{\alpha}^{-1}(\tilde{\varepsilon}).$
	\end{proposition}
	\section{Compositionality Result}\label{1:III}
	\label{s:inter}
	In this section, we analyze networks of discrete-time switched subsystems and leverage sufficient small-gain type conditions under which one can construct an alternating simulation function from a network of finite abstractions to the concrete network by using alternating simulation functions of the subsystems. In the following, we define first a network of discrete-time switched subsystems.
	\subsection{Interconnected Systems}
	We consider $N\in\N_{\ge1}$ discrete-time switched subsystems $$\Sigma_i=(\mathbb X_i,P_i,\mathbb W_i,F_i,\mathbb Y_i,h_i),i\in[1;N],$$ with partitioned internal inputs as
	\begin{small}
		\begin{align}\label{eq:int1}
		w_i\!\!=\!\![w_{i1};\ldots;w_{i(i-1)};w_{i(i+\!1)};\ldots;w_{iN}],
		\mathbb{W}_i\!\!=\!\!\prod_{j=1}^{N-1} \!\mathbb{W}_{ij},
		\end{align}
	\end{small}
	and with output map and set partitioned as
	\begin{small}
		\begin{align}\label{eq:int2}
		h_{i}(x_i)=[h_{i1}(x_i);\ldots; h_{iN}(x_i)],
		\mathbb Y_i=\prod_{j=1}^N  \mathbb Y_{ij}.
		\end{align}
	\end{small}
	$\!\!$We interpret the outputs $y_{ii}$ as external ones, whereas $y_{ij}$ with $i\neq j$ are internal ones which are used to define the 
	interconnected switched systems.
	In particular, we assume that $w_{ij}=y_{ji}$, if there is connection from switched subsystem $\Sigma_{j}$ to
	$\Sigma_i$, otherwise we set $h_{ji}\equiv 0$.
	Next, given input-output structure as in $\eqref{eq:int1}$ and $\eqref{eq:int2}$, we define the interconnection of switched subsystems.
	\begin{definition}\label{def:5}
		Consider $N\in\N_{\ge1}$ switched subsystems $\Sigma_i=(\mathbb X_i,P_i,\mathbb W_i,F_i,\mathbb Y_i,h_i)$, $i\in[1;N]$, with the input-output structure given
		by $\eqref{eq:int1}$ and $\eqref{eq:int2}$. The \emph{interconnected switched
			system} $\Sigma=(\mathbb X,P,F,\mathbb Y,h)$,
		denoted by
		$\mathcal{I}(\Sigma_1,\ldots,\Sigma_N)$, is defined by $\mathbb X =\prod_{i=1}^N \mathbb X_i$,
		$  P=\prod_{i=1}^N  P_i$, ${F}=\prod_{i=1}^N{F}_i$, $ \mathbb Y=\prod_{i=1}^N  \mathbb Y_{ii}$, 
		and map
		$h(x)\!\Let \!\intcc{h_{11}(x_1);\ldots;h_{NN}(x_N)}$, where $x=\intcc{x_{1};\ldots;x_{N}}$, and subject to the constraint:
		\begin{align}\label{const}
		\forall i,j\!\in\![1;N],i\neq j,w_{ij}=y_{ji},{\mathbb{Y}}_{ji}\!\subseteq\! \mathbb{W}_{ij}.
		\end{align}
		Similarly, given transition subsystem $T_i(\Sigma_i),i\in[1;N]$, one can also define the network of those transition subsystems as $\mathcal{I}(T_1(\Sigma_1),\ldots,T_N(\Sigma_N))$. 
	\end{definition}
	
	Next subsection provides one of the main results of the paper on the compositional construction of  abstractions for networks of switched systems. 
	\subsection{Compositional Abstractions of
		Interconnected Switched Systems}
	In this subsection, we assume that we are given $N$ discrete-time switched subsystems $\Sigma_i=(\mathbb X_i,P_i,\mathbb W_i,F_i,\mathbb Y_i,h_i),i\in[1;N]$, or equivalently, $T_i(\Sigma_i)=(X_i,U_i,W_i,\TF_i,Y_i,\mathcal H_i),$ together with their corresponding  abstractions
	$\hat{T}_i(\hat{\Sigma}_i)=(\hat{X}_i,\hat{U}_i,\hat{W}_i,\hat{\TF}_i,\hat{Y}_i,\hat{\mathcal H}_i)$ and alternating simulation functions $\SF_i$ from
	$\hat{T}_i(\hat{\Sigma}_i)$ to $T_i(\Sigma_i)$. Moreover, for $\sigma_i$, $\alpha_i$, and $\varrho_i$ associated with $\SF_i$, $\forall~ i\in [1;N]$, appeared in Definition \ref{sf}, we define
	\begin{small}
		\begin{align}\label{gammad}
		\!\!\gamma_{ij}(s)\!\!\Let\!\!\left\{
		\begin{array}{lr}
		\!\!\!\sigma_{i}s\quad\quad\quad\quad~~ \text{if}\quad i=j,\\
		\!\!\!\varrho_i\circ\alpha_{j}^{-1}(s)\quad\, \text{if} \quad i\neq j, 
		\end{array}\right. \forall s\!\in\! \R_{\ge0},\forall i,j \!\in\!\! [1;N].
		\end{align}
	\end{small}
	
	
	We raise the next small-gain assumption to establish
	the main compositionality results of the paper.
	\begin{assumption}\label{sg}
		Assume that functions $\gamma_{ij}$ defined in \eqref{gammad} satisfy
		\begin{small}
			\begin{align}\label{SGC}
			\gamma_{i_1i_2}\circ\gamma_{i_2i_3}\circ\cdots\circ\gamma_{i_{r-1}i_r}\circ\gamma_{i_ri_1}<\mathcal{I}_d,
			\end{align}
		\end{small}
		$\forall(i_1,\ldots,i_r)\in\{1,\ldots,N\}^r$, where $r\in \{1,\ldots,N\}$.\\
	\end{assumption}
	\vspace{-0.2cm}
	The next theorem provides a compositional approach on the construction of abstractions of networks of discrete-time switched subsystems and that of the corresponding alternating simulation functions. 
	\begin{theorem}\label{thm:3}
		Consider the interconnected transition system
		$T(\Sigma)=(X,U,\TF,Y,\op)$ induced by
		$N\in\N_{\ge1}$
		transition subsystems~$T_i(\Sigma_i),\forall~ i\in [1;N]$. Assume that each $T_i(\Sigma_i)$ and its abstraction $\hat{T}_i(\hat{\Sigma}_i)$ admit an alternating simulation function $\SF_i$ as in Definition \ref{sf}.
		Let Assumption \ref{sg} hold.
		Then, there exist $\delta_i \in \mathcal{K}_{\infty}$ such that 
		\begin{small}
			\begin{align}\notag
			\tilde{\SF}&((x,p,l),(\hat{x},p,l))\Let\max\limits_{i\in[1;N]}\{ \delta^{-1}_{i}\circ \SF_i((x_i,p_i,l_i),(\hat{x}_i,p_i,l_i)) \} 
			\end{align}
		\end{small}
		$\!\!\!$is an alternating simulation function from $\hat{T}(\hat{\Sigma})={\mathcal{I}}(\hat{T}_1(\hat{\Sigma}_1),\ldots,\hat{T}_{N}(\hat{\Sigma}_{N}))$ to $T(\Sigma)=\mathcal{I}(T_1(\Sigma_1),\ldots,T_{N}(\Sigma_{N}))$.
	\end{theorem}
	\section{Construction of Finite Abstractions}\label{1:IV}
	In this section, we consider $\Sigma=(\mathbb X,P,\mathbb W,F,\mathbb Y,h)$ as an infinite, deterministic switched system,  and assume its output map $h$ satisfies the following general Lipschitz-like assumption: there exists an $\ell\!\in\!\KK$ such that $\Vert h(x)\!-\!h(x')\Vert\leq\!\! \ell(\Vert x-x'\Vert)$ for all $x,x'\in \mathbb X$. In addition, the existence of an alternating simulation function between $T(\Sigma)$ and its finite abstraction is established under the assumption that $\Sigma_p$ is incrementally input-to-state stable ($\delta$-ISS) \cite{ruffer} as defined next.
	\begin{definition}\label{def:SFD1} 
		System $\Sigma_{p}$  is $\delta$-ISS if there exist functions $ V_p:\mathbb X\times \mathbb X \to \mathbb{R}_{\geq0} $, $\underline{\alpha}_p, \overline{\alpha}_p, \rho_{p} \in \mathcal{K}_{\infty}$, and constant $0<\kappa_p<1$, such that for all $x,\hat x\in \mathbb{X}$, and for all $w,\hat w\in \mathbb{W}$  
		\begin{small}
			\begin{align}\label{e:SFC11}
			\underline{\alpha}_p (\Vert x-\hat{x}\Vert ) \leq V_p(x,\hat{x})\leq \overline{\alpha}_p (\Vert x-\hat{x}\Vert ),
			\end{align}
			\begin{align}\label{e:SFC22}
			V_p(f_p(x,w),f_p(\hat x,\hat w))
			\!\leq\! \kappa_p V_p(x,\hat{x})+\rho_{p}(\Vert w- \hat{w}\Vert ).\\\notag
			\end{align}
		\end{small}
	\end{definition}
	\vspace{-0.2cm}
	We say that $V_p$, $\forall p\in P$, are multiple $\delta$-ISS Lyapunov functions for system $\Sigma$ if it satisfies \eqref{e:SFC11} and \eqref{e:SFC22}. Moreover, if $V_p=V_{p'}, \forall p,p'\in P$, we omit the index $p$ in \eqref{e:SFC11}, \eqref{e:SFC22}, and say that $V$ is a common $\delta$-ISS Lyapunov function for system $\Sigma$. We refer interested readers to \cite{liberzon} for more details on common and multiple Lyapunov functions for switched systems.  
	
	Now, we show how to construct a finite abstraction $\hat T(\hat{\Sigma})$ of transition system $T(\Sigma)$ associated to the switched system $\Sigma$ in which $\Sigma_{p}$  is $\delta$-ISS.
	\begin{definition}\label{smm} Consider a transition system $T(\Sigma)=(X,U,W,\TF,Y,\op)$, associated to the switched system $\Sigma=(\mathbb X,P,\mathbb W,F,\mathbb Y,h)$, where $\mathbb X,\mathbb W$ are assumed to be finite unions of boxes. Let $\Sigma_{p}$ be $\delta$-ISS as in Definition \ref{def:SFD1}. Then one can construct a finite transition system $\hat{T}(\hat{\Sigma})=(\hat{X},\hat{U},\hat{W},\hat{\TF},\hat{Y},\hat{\op})$ where:
		\begin{itemize}
			\item $\hat{X}=\hat{\mathbb{X}}\times P\times \{0,\cdots,k_d-1\}$, where $\hat{\mathbb{X}}=[\mathbb{X}]_{\eta}$ and $0<\eta\leq\emph{span}(\mathbb{X})$ is the state set quantization parameter; 
			\item $\hat{U}=U=P$ is the external input set;
			\item $\hat{W}=[\mathbb{W}]_{{\varpi}}$, where $0\leq{{\varpi}}\leq\emph{span}(\mathbb W)$ is the internal input set quantization parameter.
			\item $(\hat{x}',p',l')\in \hat{\TF}((\hat{x},p,l),\hat{u},\hat{w})$ if and only if $\hat{x}'\in \hat{f}_p(\hat{x},\hat{w})\Leftrightarrow\Vert f_p(\hat{x},\hat{w})-\hat{x}'\Vert\leq \eta$, $\hat{u}=p$ and the following scenarios hold:
			\begin{itemize}
				\item $l<k_d-1$, $p'=p$ and $l'=l+1$;
				\item $l=k_d-1$, $p'=p$ and $l'=k_d-1$;
				\item $l=k_d-1$, $p'\neq p$ and $l'=0$;
			\end{itemize}
			\item $\hat{Y}=\{\op(\hat{x},p,l)|(\hat{x},p,l)\in \hat{X}\}$;
			\item $\hat{\op}:\hat{X}\rightarrow \hat{Y}$ is the output map defined as $\hat{\op}(\hat{x},p,l)={\op}(\hat{x},p,l)=h(\hat{x})$;
		\end{itemize} 
	\end{definition}
	
	\begin{remark} In the context of networks of subsystems, $\hat{W}$ should be constructed 
			in such a way that it satisfies \eqref{eq:int1} and \eqref{const} in the compositional setting with respect to outputs sets of other finite transition subsystems.
	\end{remark}
	We impose the following assumptions on function $V_p$ in Definition \ref{def:SFD1} which are used to prove some of the main results later. 
	\begin{assumption}\label{ass1} 
		There exists $\mu \geq 1$ such that
		\begin{small}
			\begin{align}\label{mue}
			\forall x,y \in \mathbb{X},~~ \forall p,p' \in P,~~ V_p(x,y)\leq \mu  V_{p'}(x,y). 
			\end{align}
		\end{small}
	\end{assumption}
	\vspace{0.1cm}
	\begin{assumption}\label{ass2} 
		For all $p\in P$, there exists a $\mathcal{K}_{\infty}$ function $\gamma_p$ such that
		\begin{small}
			\begin{align}\label{tinq} 
			\forall x,y,z \in \mathbb{X},~~V_p(x,y)\leq V_p(x,z)+\gamma_p(\Vert y-z\Vert). 
			\end{align}
		\end{small}
	\end{assumption}
	\vspace{0.1cm}
	
	Now, we establish the relation between $T(\Sigma)$ and $\hat{T}(\hat{\Sigma})$, introduced above, via the notion of alternating simulation function as in Definition \ref{sf}.
	\begin{theorem}\label{thm1}
		Consider a switched system $\Sigma=(\mathbb X,P,\mathbb W,F,\mathbb Y,h)$ with its equivalent transition system $T(\Sigma)=(X,U,W,\TF,Y,\op)$. Let $\Sigma_{p}$ be $\delta$-ISS as in Definition \ref{def:SFD1}. Consider a finite transition system $\hat{T}(\hat{\Sigma})=(\hat{X},\hat{U},\hat{W},\hat{\TF},\hat{Y},\hat{\op})$ constructed as in Definition \ref{smm}. Assume that Assumptions \ref{ass1} and \ref{ass2} hold. 	
		Let $\epsilon>1$. If, $\forall p \in P, ~k_d\geq \epsilon \frac{\ln (\mu)}{\ln (\frac{1}{\kappa_p})}+1$, then function $\mathcal{V}$ defined as 
		\begin{small}
			\begin{align}\label{sm}
			\mathcal{V}((x,p,l),(\hat{x},p,l))\Let {\kappa^{\frac{-l}{\epsilon}}_pV_p(x,\hat{x})}{}
			\end{align}
		\end{small}
		$\!$is an alternating simulation function from $\hat{T}(\hat{\Sigma})$ to $T(\Sigma)$. 
	\end{theorem}
	\begin{remark}\label{nonl-common}
		If $\Sigma$ admits a common $\delta$-ISS Lyapunov function satisfying Assumption \ref{ass2}, then function $\mathcal{V}$ in Theorem \ref{thm1} reduces to $\mathcal{V}((x,p,l),(\hat{x},p,l))\Let V(x,\hat{x})$.
	\end{remark}
	\section{Case Study}\label{1:V}
	The chosen switched system $\Sigma$ here is the
	model of a circular road around a city (Highway) divided into $50$ cells of $1000$ meters each. The road has
	$25$ entries and $50$ ways out in such a way that cell $q$ has an entry and exit if $q\in Q_1=\{q ~\text{is odd}~ | q\in [1;50]\}$ and has an exit and no entry if $q\in Q_2=\{q~\text{is even}~ | q\in [1;50]\}$. The entries are controlled by
	traffic signals, denoted $s_r, r\in[1;25]$, that enable (green light) or
	not (red light) the vehicles to pass. 
	In $\Sigma$, the dynamic we want to observe is the density of traffic, given in vehicles per cell, for each cell $q$
	of the road. The state of switched system $\Sigma$ is a $50$-dimensional vector and its set of modes can be understood as all possible linear combination of traffic signals $s_r$. More formally, since each traffic signal $s_r$ can have two modes ($1$ for red light and $2$ for green), one can consider the modes of system $\Sigma$ as $p\in P=\{1,2\}^{25}$.
	\begin{figure}
		\begin{center}
			\includegraphics[height=7cm]{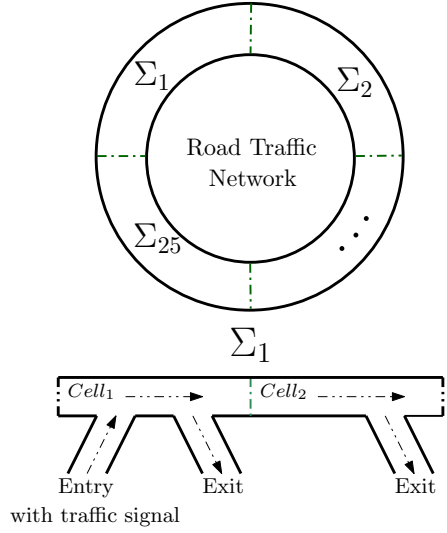}
			\caption{{\small Model of a road traffic network in a circular highway composed of 25 identical links, each link has two cell.}}
			\label{rdn}
		\end{center}
		\vspace*{-0.7cm}
	\end{figure}
	During the sampling time interval $\tau=\frac{10}{60\times60}$ in hours ($h$), we assume that $12$ vehicles can pass the entry controlled by a traffic signal $s_r$ when it is green. Moreover, $10\%$ of vehicles that are in cells $q\in Q_1$, and $35\%$ of vehicles that are in cells $q\in Q_2$ go out using available exits. As explained in \cite{Kibangou}, the evolution of the density $\mathbf{x}$ of all cells are described by the interconnected discrete-time switched model:
		\begin{align*}
		\Sigma:\left\{
		\begin{array}{rl}
		\mathbf{x}(k+1)&=A\mathbf{x}(k)+ B_{\mathsf{p}(k)},\\
		\mathbf{y}(k)&=\mathbf{x}(k),
		\end{array}\right.
		\end{align*}
	$\!\!\!\!$where $A\in\R^{50\times 50}$ is a matrix with elements $\{A\}_{q,q}=0.9-\frac{\tau v}{d}$ if $q\in Q_1$ and $\{A\}_{q,q}=0.65-\frac{\tau v}{d}$ if $q\in Q_2$, $\{A\}_{q+1,q}=\{A\}_{1,50}=\frac{\tau v}{d}$, $\forall q\in [1;50]$, and all other elements are identically zero,
	where $d=1$ and $v=120$ are the length in kilometers ($km$) and the flow
	speed of the vehicles in kilometers per hour ($km/h$), respectively. 
	The vector $B_p\in \R^{50}$ is defined as ${B}_p=\intcc{b_{1p_1};\ldots; b_{25p_{25}}}$ such that $b_{ip_i}=[0;0]$ if $p_i=1$, and 
	$b_{ip_i}=[0;12]$ if $p_i=2$, $\forall i\in[1,25]$, $\intcc{p_{1};\ldots;p_{25}}\in P=\{1,2\}^{25}$, where $P$ is the set of modes of $\Sigma$. 
	Now, in order to apply the compositionality result, we introduce subsystems $\Sigma_i$, $\forall i \in [1;25]$. Each subsystems $\Sigma_i$ represents the dynamic of one link of the entire highway, where each link contains $2$ cells, one entry, and two exits as illustrated in Fig \ref{rdn}. The subsystems $\Sigma_i$ is described by
		\begin{align*}
		\Sigma_i:\left\{
		\begin{array}{rl}
		\mathbf{x}_i(k+1)&=A_i\mathbf{x}_i(k)+ D_{i}w_i(k)+B_{i\mathsf{p}_i(k)},\\
		\mathbf{y}_i(k)&=C_i\mathbf{x}_i(k),
		\end{array}\right.
		\end{align*}
	where, $\forall i\in[1;25]$, 
		\begin{align*}
		&A_i\!=\!\begin{bmatrix}
		0.9-\frac{\tau v}{d} & 0  \\
		\frac{\tau v}{d} & 0.65-\frac{\tau v}{d}\\
		\end{bmatrix}\!,
		D_i\!=\!\begin{bmatrix}
		\frac{\tau v}{d} \\
		0\\
		\end{bmatrix}\!,
		B_{i1}\!=\!\begin{bmatrix}
		0  \\
		0\\
		\end{bmatrix}\!,\\
		&B_{i2}\!=\!\begin{bmatrix}
		12  \\
		0\\
		\end{bmatrix}\!,
		C_{i}\!=\!\begin{bmatrix}
		C_{ii}\\
		C_{i(i+1)}\\
		\end{bmatrix}\!,
		C_{ii}\!=\!\begin{bmatrix}
		1& 0\\
		0& 1
		\end{bmatrix}\!,
		C_{i(i+1)}\!=\!\begin{bmatrix}
		0& 1
		\end{bmatrix}\!,
		\end{align*}
	$\!\!\omega_i(k)=C_{(i-1)i}\mathbf{x}_{i-1}(k)$ (with $C_{01}\Let C_{N(N+1)}$, and $\mathbf{x}_{0}\Let\mathbf{x}_{N},N=25$), and the set of modes is $P_i=\{1,2\}$. 
	Clearly, one can verify that $\Sigma=\mathcal{I}(\Sigma_1,\ldots,\Sigma_{25})$.
	
	$\!\!\!\!\!\!\!$Note that, for any $i\in[1;25]$,
	conditions \eqref{e:SFC11} and \eqref{e:SFC22} are satisfied with $V_{ip_i}(x_i,\hat{x}_i)= \Vert x_i-\hat x_i\Vert$, $\underline{\alpha}_{ip_i}=\overline{\alpha}_{ip_i}=\mathcal{I}_d$, $\kappa_{ip_i}= 0.65$, $\rho_{ip_i}=0.33\mathcal{I}_d$, $\forall p_i\in P_i$.  
	Furthermore, condition \eqref{tinq} is satisfied with $\gamma_{ip_i}=\mathcal I_d$, $\forall p_i\in P_i$. Moreover, since $V_{ip_i}= V_{ip'_i},\forall p_i,p_i'\in P_i$, and according to Remark \ref{nonl-common}, function $\mathcal{V}_i((x_i,p_i,l_i),(\hat{x}_i,p_i,l_i))= \Vert x_i-\hat x_i\Vert$ is an alternating simulation function from $\hat{T}_i(\hat{\Sigma}_i)$ to $T_i(\Sigma_i)$
	Note that for the construction of finite abstractions, we have chosen the finite set $\hat{{W}}_{i}=\{C_{(i-1)i}\hat{x}_{i-1}|\hat{x}_{i-1}\in \hat{\mathbb{X}}_{i-1}\},~\forall i\!\in\![1;25]$, (with $C_{01}\Let C_{N(N+1)}$, $\hat{x}_{0}\Let\hat{x}_{N}$, and $\hat{\mathbb{X}}_{0}\Let\hat{\mathbb{X}}_{N}, N=25$).
	Now, by employing \eqref{gammad}, we have $\gamma_{ij}<\mathcal{I}_d$, $\forall i,j\in[1;25]$, hence the small-gain condition \eqref{SGC} is satisfied.
	Using the results in Theorem \ref{thm:3} with $\delta^{-1}_{i}=\mathcal{I}_d,~ \forall i\in [1;25]$, one can verify that $\mathcal{V}((x,p,l),(\hat{x},p,l))\!=~\!\!\max_{i}\{\!\Vert x_i-\hat x_i\Vert\}$ is an alternating simulation function from ${\mathcal{I}}(\hat{T}_1(\hat{\Sigma}_1),\ldots,\hat{T}_{25}(\hat{\Sigma}_{25}))$ to $\mathcal{I}(T_1(\Sigma_1),\ldots,T_{25}(\Sigma_{25}))$.
	
\begin{figure}
		\begin{center}
			\includegraphics[height=6cm, width=10cm]{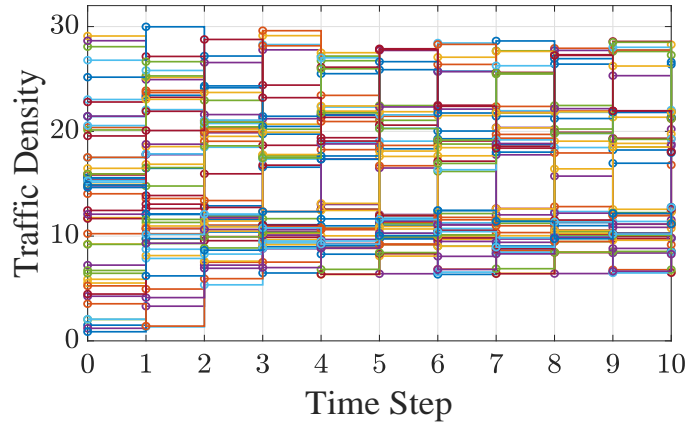}
			\caption{{\small Closed-loop state trajectories of $\Sigma$ consisting of $50$ cells.}}
			\label{st1}
		\end{center}
		\vspace*{-0.7cm}
	\end{figure}

	Next we design a controller for $\Sigma$ via finite abstractions $\hat{T}_i(\hat{\Sigma}_i)$ such that the controller maintains
	the density of traffic lower than $30$ vehicles per cell. The idea here is to design local controllers
	for finite abstractions $\hat{T}_i(\hat{\Sigma}_i)$, and then use them in concrete switched subsystems $\Sigma_i$. To do so, the local controllers are designed
	while assuming that the other subsystems meet their specifications. The computation times for constructing abstractions and designing controllers for $\Sigma_i$ with state quantization parameter $\eta_i=0.03$ are $10.2s$ and $0.014s$, respectively.
	Figure \ref{st1} shows the closed-loop state trajectories of the of $\Sigma$ consisting of $50$ cells. Note that  it would not have been possible to synthesize a controller for the $50$-dimensional switched system $\Sigma$  without applying the proposed compositional method.

	\bibliographystyle{ieeetr}
	\bibliography{ref}         
	
	
	
		\appendix
	\subsection{Notation}
	We denote by $\R$, $\Z$, and $\N$ the set of real numbers, integers, and non-negative integers,  respectively.
	These symbols are annotated with subscripts to restrict them in
	the obvious way, e.g., $\R_{>0}$ denotes the positive real numbers.
	Given $N\in\N_{\ge1}$, vectors $\nu_i\in\R^{n_i}$, $n_i\in\N_{\ge1}$, and $i\in[1;N]$, we
	use $\nu=[\nu_1;\ldots;\nu_N]$ to denote the vector in $\R^n$ with
	$n=\sum_i n_i$ consisting of the concatenation of vectors~$\nu_i$.
	The closed interval in $\N$ is denoted by $[a;b]$ for $a,b\in\N$ and $a\le b$. 
	We denote by $\mathsf{diag}(A_1,\ldots,A_N)$ the block diagonal matrix with diagonal matrix entries $A_1,\ldots,A_N$.
	We denote the identity matrix in $\R^{n\times n}$ by $I_n$. 
	The individual elements in a matrix $A\in \R^{m\times n}$, are denoted by $\{A\}_{ij}$, where $i\in[1;m]$ and $j\in[1;n]$. We denote by $\norm{\cdot}$ the infinity norm. 
	We denote by $|\cdot|$ the cardinality of a given set and by $\emptyset$ the empty set. For any set \mbox{$S\subseteq\R^n$} of the form of finite union of boxes, e.g., $S=\bigcup_{j=1}^MS_j$ for some $M\in\N$, where $S_j=\prod_{i=1}^n [c_i^j,d_i^j]\subseteq \R^n$ with $c^j_i<d^j_i$, and positive constant $\eta\leq\emph{span}(S)$, where $\emph{span}(S)=\min_{j=1,\ldots,M}\eta_{S_j}$ and \mbox{$\eta_{S_j}=\min\{|d_1^j-c_1^j|,\ldots,|d_n^j-c_n^j|\}$}, we define \mbox{$[S]_{\eta}=\{a\in S\,\,|\,\,a_{i}=k_{i}\eta,k_{i}\in\mathbb{Z},i=1,\ldots,n\}$}.
	The set $[S]_{\eta}$ will be used as a finite approximation of the set $S$ with precision $\eta$. Note that $[S]_{\eta}\neq\emptyset$ for any $\eta\leq\emph{span}(S)$.   
	We use notations $\mathcal{K}$ and $\mathcal{K}_\infty$
	to denote different classes of comparison functions, as follows:
	$\mathcal{K}=\{\alpha:\mathbb{R}_{\geq 0} \rightarrow \mathbb{R}_{\geq 0} |$ $ \alpha$ is continuous, strictly increasing, and $\alpha(0)=0\}$; $\mathcal{K}_\infty=\{\alpha \in \mathcal{K} |$ $ \lim\limits_{r \rightarrow \infty} \alpha(r)=\infty\}$.
	For $\alpha,\gamma \in \mathcal{K}_{\infty}$ we write $\alpha<\gamma$ if $\alpha(s)<\gamma(s)$ for all $s>0$, and $\mathcal{I}_d\in\mathcal{K}_{\infty}$ denotes the identity function.   
\end{document}